\documentclass{mem}
\usepackage{natbib}\usepackage{txfonts}\usepackage{balance}
\usepackage{flushend}
\usepackage{graphicx}
\usepackage{multirow}
\usepackage{xcolor}
\usepackage{ulem}
\usepackage[breaklinks,pdftex]{hyperref}
\idline{75}{282}
\begin{document}
\def\teff{$T\rm_{eff }$}
\def\kms{$\mathrm {km s}^{-1}$}

\newcommand{\EA}[1]{\textcolor[rgb]{0.5, 0., 0.5}{#1}}
\newcommand{\EAQ}[1]{\textcolor[rgb]{0., 0.5, 0.}{#1}}
\title{Massive star clusters in the gamma-ray sky

 }

   \subtitle{The role of H{\large II} regions}

\author{
G. \,Peron\inst{1} 
\and  G. \, Morlino\inst{1} \and S. \, Gabici \inst{2} \and  E. \, Amato \inst{1} }

\institute{$^{1}$ Istituto Nazionale di Astrofisica --
Osservatorio Astrofisico di Arcetri, Largo Enrico Fermi 5, 50125, Firenze, Italy \\ $^{2}$ Université de Paris, CNRS, Astroparticule et Cosmologie  F-75013 Paris, France \\
\email{giada.peron@inaf.it}\\
}

\authorrunning{G. Peron}

\titlerunning{Gamma-rays from H\textsc{ii} regions}

\date{Received: XX-XX-XXXX (Day-Month-Year); Accepted: XX-XX-XXXX (Day-Month-Year)}

\abstract{Massive Star Clusters (SCs) have been proposed as important CR sources, with the potential of explaining the high-energy end of the Galactic cosmic-ray (CR) spectrum, that Supernova Remnants (SNRs) seem unable to account for.
Thanks to fast mass losses due to the collective stellar winds, the environment around SCs is potentially suitable for particle acceleration up to PeV energies and the energetics is enough to account for a large fraction of the Galactic CRs, if the system is efficient enough. A handful of star clusters have been detected in gamma-rays confirming the idea that particle acceleration is taking place in this environment. However, contamination by other sources often makes it difficult to constrain the contribution arising from SCs only.
Here we present a new analysis of Fermi-LAT data collected towards a few massive young star clusters. The young age ($<$ 3 Myr) of the clusters guarantees that no SN has exploded in the region, allowing us to determine the power contributed by the stellar component alone, and to quantify the contribution of this type of sources to the bulk of CRs. Moreover, we will present a recent statistical investigation that quantifies the degree of correlation between gamma-ray sources and these astrophysical objects and briefly discuss the observational prospect for ASTRI and CTAO.

}
\maketitle{}

\section{Introduction}
Young massive star clusters (YMSCs) are characterized by powerful winds \citep{Seo2018TheProduction,Celli2024Mass}, potentially able to accelerate particles \citep{GabiciICRC,Morlino2021ParticleClusters}. If efficient conversion from wind luminosity to particle acceleration takes place, star clusters should emerge in gamma-rays. Regions of star formation have been extensively detected in gamma-rays from GeV to PeV energies, but a clear association of the emission with still active massive stars, rather than with their after-life products such as pulsars (PSRs) and supernova remnants (SNRs), has been possible only in a few cases. The limited angular resolution of gamma-ray instruments often fails in disentangling crowded regions, where several potential counterparts are found. The only way to overcome such source confusion, and determine the amount of accelerated particles provided by stellar winds is to target very young systems ($\lesssim$3 Myr), where no supernova has exploded yet. In the earliest phases of their life, YMSCs are found still in the gas cocoon out of which they form, often detected as an H\textsc{ii} region, and where star formation is still ongoing. This means that for a fraction of time stars blow winds and potentially accelerate particles in a rather dense environment, with density of the order of 10$^2$--$10^3$ cm$^{-3}$, serving as a perfect target for hadronic interactions and consequent gamma-ray production.  Emission from the HII regions RCW~38, RCW~36 and RCW~32 emerged in Fermi-LAT by analyzing the region of the Vela Molecular cloud Ridge (VMR) \citep{Peron2024ThePopulation}. The spectral shape and the spatial coincidence with dense gas support the hadronic origin of the emission and allow one to constrain particle acceleration. 

\section{Correlation with gamma-ray sources}
\begin{figure}[h!]
    \centering
    \includegraphics[width=1\linewidth]{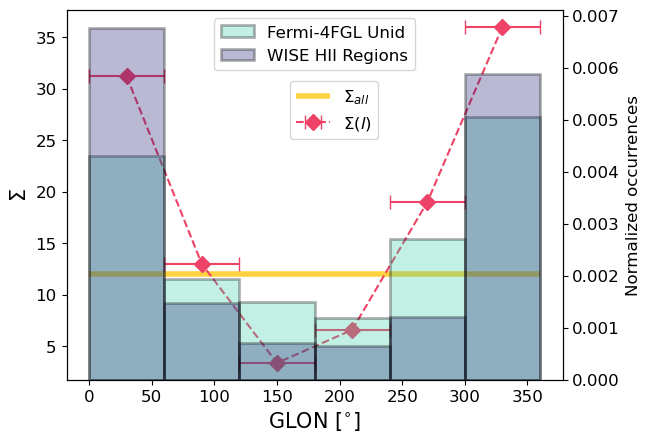}
    \caption{Statistical significance of associations of H\textsc{ii} regions with Fermi-LAT unidentified sources, as a function of their galactic longitude (red errorbars). The histogram underneath shows the distribution in longitude of the considered H\textsc{ii} regions \citep{Anderson2014TheRegions} and Fermi-LAT unidentified sources \citep{Abdollahi2022IncrementalCatalog}. Both distributions are normalized.}
    \label{fig:sig_l}
\end{figure}

The detected H\textsc{ii} regions analyzed in \cite{Peron2024ThePopulation} were found to be coincident with unidentified Fermi-LAT sources. We searched then whether a significant correlation emerged beyond the Vela region by correlating unidentified sources of the Fermi-LAT catalog with the WISE catalog of H\textsc{ii} regions \citep{Peron2024Correlation}. The latter has been constructed with WISE infrared data collected at 22~$\mu$m, which trace the emission of hot dust around stars \cite{Anderson2014TheRegions}. The association was made considering the geometrical distance between the sources: the two counterparts are considered as \textquotedblleft matching" if they overlap geometrically within the radius of the H\textsc{ii} region. The significance of the association was then evaluated by simulating 1000 catalogs of sources with random extraction in $l$ and $b$, and by comparing the number of matches obtained with the real catalogs to the average number of matching obtained with the simulated catalogs. The 
significance of the correlation was defined as 
$$ \Sigma = \frac{N_{\rm real}-<N_{\rm sim}>}{\sigma_{\rm sim}} \,,$$ 
with $N_{\rm sim}$ and $\sigma_{\rm sim}$ the average and the standard deviation of the number of matches with the simulated catalogs. {The correlation was found to be highly significant: we found 127 H\textsc{ii} regions overlapping with at least one unidentified Fermi source, compared to the average in the simulations of $\sim$ 50 matches; this corresponds to $\Sigma=12$.} 
We also tested a possible dependence of the correlation on the position of the H\textsc{ii} regions along the plane. We repeated the test by considering slices of 60$^{\circ}$ in galactic longitude. The results are shown in Figure \ref{fig:sig_l}. As one can see, the significance is higher in the inner part of the Galaxy, {with $\Sigma>35$, in spite of the fact that a larger number of sources are naturally found in the inner Galactic regions, potentially worsening source confusion.}

{On the other hand, the projected size of the H\textsc{ii} regions is larger in the outskirts of the disk, and this could in principle increase the chance coincidence. We note also that the distribution of H\textsc{ii} regions declines more rapidly towards the outer part of the Galaxy than the Fermi sources, suggesting a different interpretation for the counterparts of these sources in this region. }


\section{Fermi-LAT observations }
\begin{figure*}
    \centering
    \includegraphics[width=0.45\linewidth]{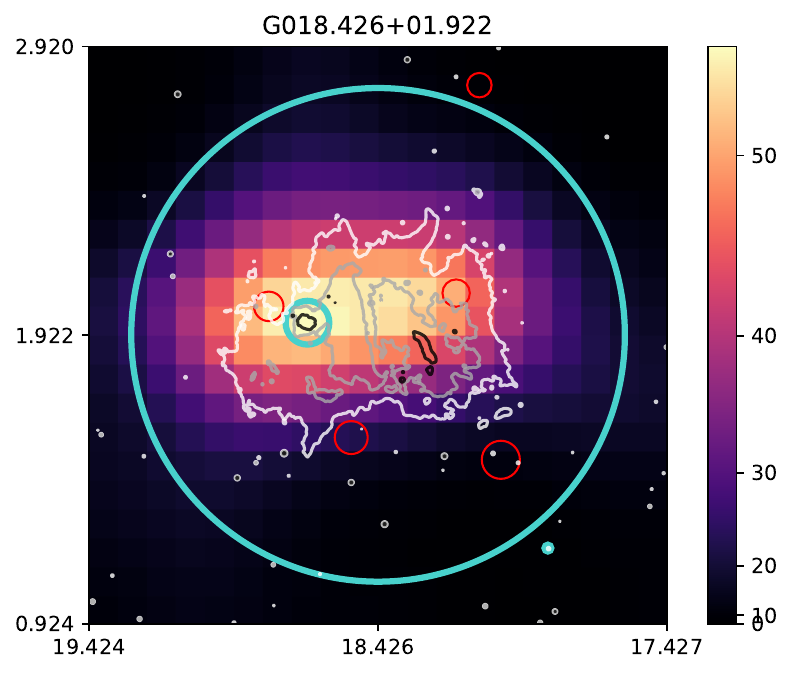}\includegraphics[width=0.45\linewidth]{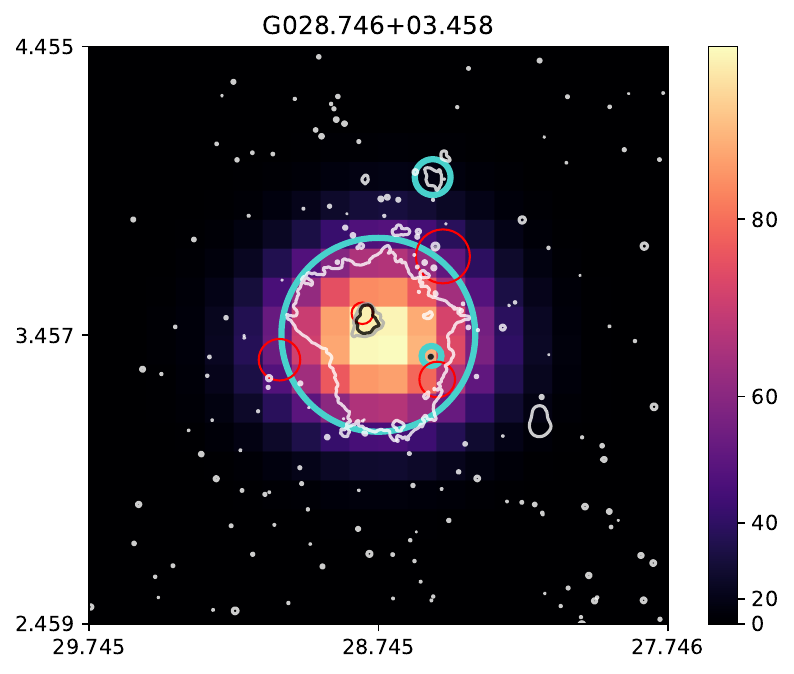}
    \includegraphics[width=0.45\linewidth]{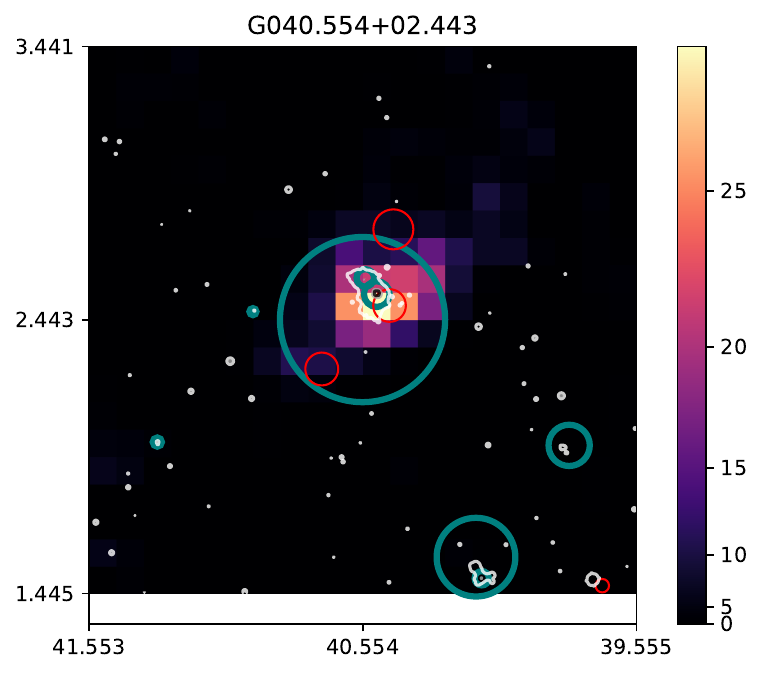}\includegraphics[width=0.45\linewidth]{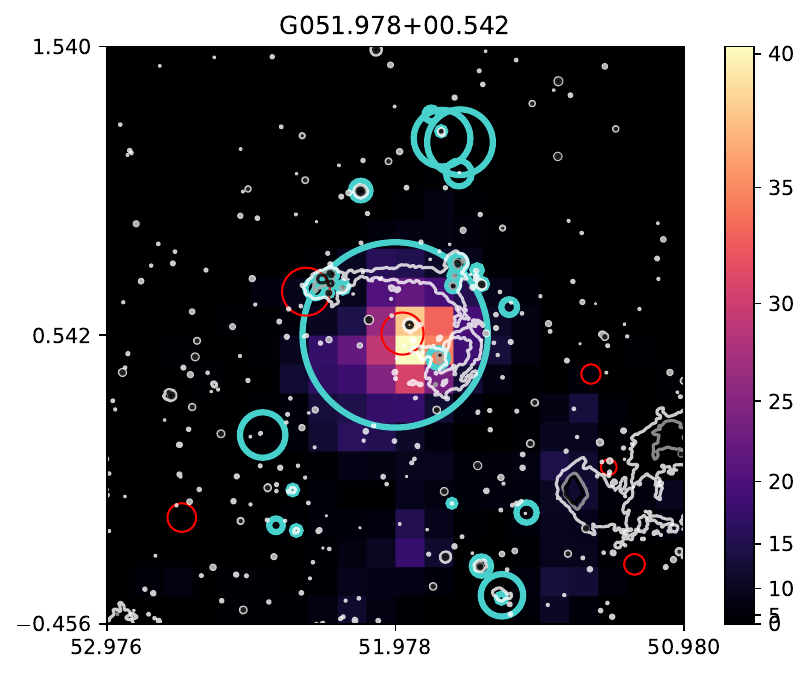}
    \caption{Test statistic maps obtained from Fermi-LAT data after the optimization of the background sources and the removal of the  sources potentially associated to  HII regions. The light-blue circles represent the extension of the HII regions in the field, while the contours represent their 22-$\mu$m emission as traced by WISE.}
    \label{fig:tsmaps}
\end{figure*}

\begin{figure*}
    \centering
    \includegraphics[width=0.49 \linewidth]{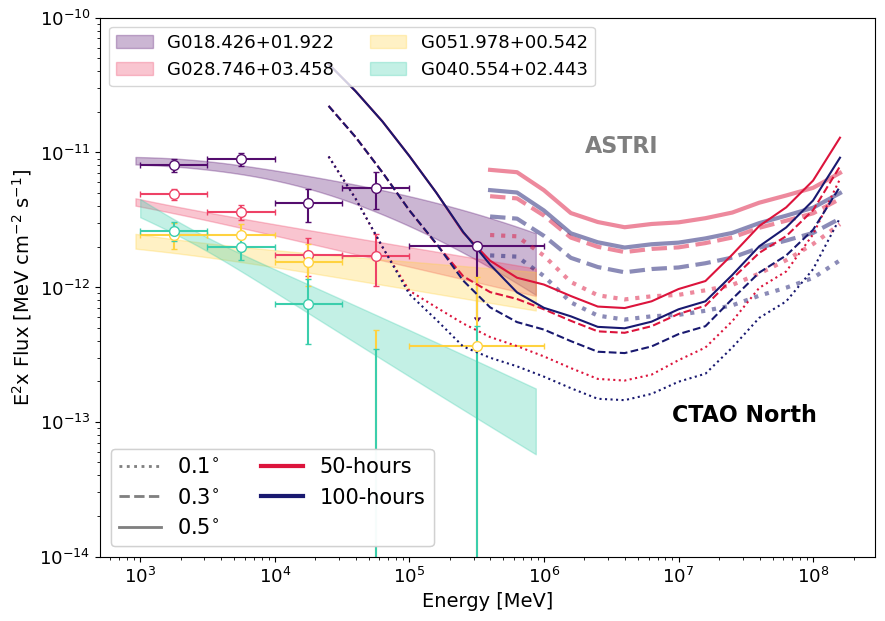}\includegraphics[width=0.49 \linewidth]{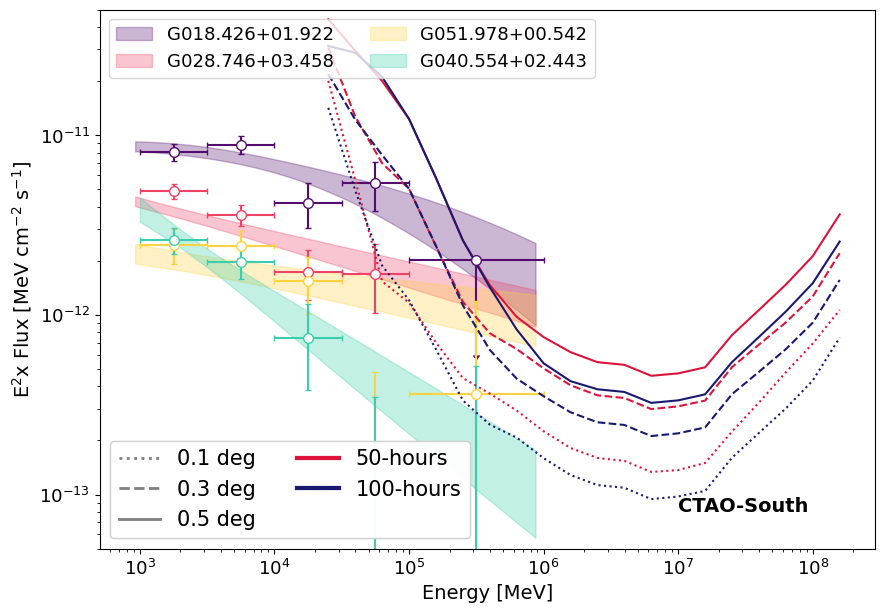}
    \caption{Spectral energy distribution of the analyzed H\textsc{ii} regions compared to the sensitivity towards extended sources of the future ASTRI-mini array and CTAO facilities. The sensitivity curves are calculated as in \cite{Celli2024DetectionLHAASO} }
    \label{fig:seds}
\end{figure*}

Among the H\textsc{ii} regions with overlapping Fermi sources, we selected a few for follow-up analysis: {the brightest, in terms of infrared flux, among the ones with size larger than the Fermi Point Spread Function ($\sim$0.2$^\circ$ above a few GeV).}
The list of targets is reported in Table \ref{tab:selection}, where we also report the Fermi sources that overlap. We performed a standard Fermi-LAT analysis where the background model is composed of Galactic and extragalactic diffuse emission as given by the Collaboration\footnote{See \href{https://fermi.gsfc.nasa.gov/ssc/data/access/lat/BackgroundModels.html}{Fermi-LAT webpage}}, and by the point sources of the Fermi-LAT 4th source catalog (4FGL \cite{Abdollahi2022IncrementalCatalog}). After optimization of the background source parameters, we removed from the model the sources that are potentially associated with the clusters as listed in Table \ref{tab:selection}: the resulting residual maps, computed in terms of test statistics (TS) are reported in Figure \ref{fig:tsmaps}. The figure shows the 4FGL sources in the region as red circles that represent the uncertainty on their localization, as reported in the catalog. The morphology of the Fermi-LAT emission matches well with the infrared emission, validating the identification of the gamma-ray emission with the H$\textsc{ii}$ regions.
We fixed the extension of our sources to the WISE extension and extracted the spectral energy distribution from them. The result is reported in Figure \ref{fig:seds}. 
The selected clusters are located near the celestial equator (DEC $\sim 0^\circ$), thus they may be accessible from ground-based observations from both the northern and the southern hemispheres. We compare the spectra with the sensitivity of the future Cherenkov telescope facilities, ASTRI\footnote{Astrofisica con Specchi a Tecnologia Replicante Italiana} and CTAO\footnote{The Cherenkov Telescope Array Observatory}, calculated for extended sources as in \cite{Celli2024DetectionLHAASO}. Even if the latter is a conservative estimate, because it requires a signal of 5$\sigma$ in each energy bin, the detected flux will be within the reach of the next generation arrays, opening the possibility of testing their emission at TeV energies.

\begin{table*}[]
    \centering
    \begin{tabular}{l|ccccccl}
\hline
WISE Name & $(l,b)^{\circ}$ & Extension [$^\circ$] & 22$\mu$m Flux [Jy] &   4FGL souces (Class) \\
\hline    
\multirow{4}{*}{G018.426+01.922} & \multirow{4}{*}{(18.426,1.923)} & \multirow{4}{*}{0.85} & \multirow{4}{*}{33317.21} &   4FGL~J1816.5-1208c (Unid)  \\
& & & &  4FGL~J1818.3-1233 (Unid)\\
& & & &  4FGL~J1817.9-1135 (Unid)\\
& & & &  4FGL~J1819.0-1203 (Bin) \\
\hline
\multirow{2}{*}{G028.746+03.458} & \multirow{2}{*}{(28.746,3.459)} & \multirow{2}{*}{0.34} & \multirow{2}{*}{7171.97} & 4FGL J1831.6-0223c (Unid) \\
 &  &  &  &   4FGL J1831.3-0203c (Unid) \\
\hline
\multirow{2}{*}{G040.554+02.443} & \multirow{2}{*}{(40.555,2.444)} & \multirow{2}{*}{0.3} & \multirow{2}{*}{572.67} &  4FGL J1857.5+0756 (Unid) \\
& & & &  4FGL~J1856.2+0749 (Unid)\\
\hline
G051.978+00.542 & (51.979,0.543) & 0.32 & 780.05 &  4FGL J1925.1+1707 (Unid)\\

\hline
    \end{tabular}
    \caption{Location, extension and IR flux of the analyzed HII regions. The last column also report the Fermi sources that overlap the regions.}
    \label{tab:selection}
\end{table*}

\section{Discussion and conclusions}
H\textsc{ii} regions are found in the surrounding of very massive young clusters, younger than the age at which SNe start to explode. Their infrared and radio emission serve as tracers for identifying clusters that are obscured at shorter wavelengths. The presence of gas, serving as a target for hadronic interactions, enhances the chances for detecting gamma-rays from accelerated particles. After the first pilot study on the Vela region \cite{Peron2024ThePopulation}, that revealed an acceleration efficiency of the order of 1 \%, we further investigated promising H\textsc{ii} regions that overlap with unidentified Fermi-LAT sources. Emission up to 1~TeV emerged from the WISE regions G018.426+01.922, G028.746+03.458, G040.554+02.443, and G051.978+00.542. The similar morphology seen in the gamma-ray and infrared emission confirms the connection between unidentified sources and the H\textsc{ii} regions. Further study would be needed to confirm whether this type of sources could be seen at higher energies. The future ASTRI-Mini Array and CTAO will help in shedding light on this issue.
 
\bibliographystyle{aa}
\bibliography{bibliography}

\end{document}